\documentclass{fairmeta}
\usepackage{amsmath}
\usepackage{enumerate} 
\usepackage{bm}
\usepackage{algorithm}
\usepackage{floatflt}
\usepackage{algpseudocode}
\usepackage{amsfonts}
\usepackage{amsthm}
\usepackage{newtxtt}
\usepackage{algorithm}
\usepackage{colortbl} 
\usepackage{cleveref}
\usepackage{diagbox} 
\usepackage[utf8]{inputenc}
\usepackage{textgreek}
\usepackage{colortbl}
\usepackage{nicematrix}
\usepackage{makecell}
\usepackage{float}
\usepackage{arydshln}
\usepackage[frozencache,cachedir=.]{minted}
\usepackage{caption}
\usepackage{subcaption}
\usepackage{tcolorbox}
\usepackage{amssymb}
\usepackage{xspace}
\usepackage{wrapfig}
\usepackage{adjustbox}
\usepackage{tabularx}
\usepackage{booktabs}
\usepackage{mathtools}
\usepackage{wrapfig}
\usepackage{amssymb}
\usepackage{graphicx}

\usepackage{silence}
\makeatletter
\patchcmd{\wrong@fontshape}{\@gobbletwo}{}{}{}
\makeatother
\definecolor{upColor}{RGB}{17,138,21}
\definecolor{downColor}{RGB}{174,36,67}

\newtheorem{theorem}{Theorem}[]

\newtheorem{remark1}[theorem]{Remark}

\title{A Real-Time Privacy-Preserving Behavior Recognition System via Edge-Cloud Collaboration}
\author[]{Huan Song}
\author[]{Shuyu Tian}
\author[]{Junyi Hao}
\author[]{Cheng Yuan}
\author[]{Zhenyu Jia}
\author[]{Jiawei Shao}
\author[]{Xuelong Li}
\affiliation[]{Institute of Artificial Intelligence (TeleAI), China Telecom}
\metadata[Correspondence to]{Xuelong Li (\email{xuelong\_li@ieee.org})}

\begin{document}

\abstract{
As intelligent sensing expands into high-privacy environments such as restrooms and changing rooms, the field faces a critical privacy-security paradox. Traditional RGB surveillance raises significant concerns regarding visual recording and storage , while existing privacy-preserving methods—ranging from physical desensitization to traditional cryptographic or obfuscation techniques—often compromise semantic understanding capabilities or fail to guarantee mathematical irreversibility against reconstruction attacks. To address these challenges, this study presents a novel privacy-preserving perception technology based on the AI Flow theoretical framework and a edge-cloud collaborative architecture. The proposed methodology integrates source desensitization with irreversible feature mapping. Leveraging Information Bottleneck theory, the edge device performs millisecond-level processing to transform raw imagery into abstract feature vectors via non-linear mapping and stochastic noise injection. This process constructs a unidirectional information flow that strips identity-sensitive attributes, rendering the reconstruction of original images impossible. Subsequently, the cloud platform utilizes multimodal family models to perform joint inference solely on these abstract vectors to detect abnormal behaviors. This approach fundamentally severs the path to privacy leakage at the architectural level, achieving a breakthrough from video surveillance to de-identified behavior perception and offering a robust solution for risk management in high-sensitivity public spaces.
}

\maketitle

\section{Introduction}
With the deepening development of smart city construction and the widespread popularity of IoT technologies, the reach of public safety governance is gradually extending into more granular scenarios \citep{an2026single,ge2023passive}. However, security management faces severe challenges in areas with high privacy sensitivity, such as restrooms, changing rooms, and hospital wards. On one hand, these concealed spaces are often hotspots for safety hazards like falls, smoking, and school bullying; on the other hand, the public holds strong psychological resistance and ethical concerns regarding being gazed at, recorded, and stored. For a long time, such places have remained blind spots in security regulation, trapping managers in a binary paradox of sacrificing privacy for safety or abandoning perception to protect privacy.

To circumvent privacy and ethical controversies associated with RGB images at the source, researchers first attempted to introduce non-visual sensors as alternatives. Naser et al. demonstrated the effectiveness of low-resolution Thermal Sensor Arrays (TSA) in scenarios such as elderly home care, pointing out their inherent de-identification advantage by capturing only heat distribution and avoiding facial texture acquisition \citep{newaz2024low}. Similarly, Ren et al. proposed a solution based on single-pixel Time-of-Flight (ToF) detection, utilizing light pulse time evolution to reconstruct 3D point clouds for action recognition, thereby thoroughly discarding the exposure of identity via color information \citep{ofodile2019action}. However, these methods relying solely on physical characteristics often face a severe "Semantic Gap." Due to the lack of texture details, such sensors struggle to accurately identify fine-grained behaviors like smoking or minor physical conflicts. To address this, the academic community introduced Event Cameras, aiming to remove background redundancy and capture high-speed motions by recording only pixel brightness changes. Nevertheless, recent research in 2022 points out that event streams are not absolute safe harbors; under specific algorithmic attacks, the retained edge information still risks reconstruction, failing to achieve absolute irreversibility in a mathematical sense.

The insurmountable gap of physical sensors in capturing fine-grained semantic information has forced the research focus to return to the RGB vision and algorithm level, which contains rich texture details, attempting to find a new equilibrium between high-fidelity perception and privacy protection through software-defined methods. In terms of visual processing, traditional Image Obfuscation is currently the most widely used means. However, a latest survey in arXiv (2025) points out that such methods face irreconcilable contradictions in balancing privacy protection and data utility; simple pixelation leads to a sharp decline in downstream task accuracy. More critically, the FaceObfuscator study published in USENIX Security (2024) proves that existing deep learning reconstruction attacks can easily penetrate protective layers like mosaics to recover original faces with high precision, rendering traditional obfuscation ineffective against strong adversarial AI; even the latest Vision-Language Models (VLMs) can infer concealed sensitive information through context. To address the risks of raw data transmission, Federated Learning (FL) and Homomorphic Encryption (HE) have garnered attention as advanced solutions. The former avoids the leakage risk of uploading video to the cloud through a "data stays, model moves" mechanism; the latter allows direct computation on ciphertext, providing strict mathematical privacy guarantees. However, evaluations of surveillance systems show that Federated Learning imposes extremely high demands on the computing power and bandwidth stability of edge nodes, while the high computational cost of Homomorphic Encryption restricts its large-scale deployment in massive camera real-time warning scenarios.

Existing single technical paths struggle to simultaneously satisfy the dual demands of high-privacy places: absolute identity unknowability and precise risk perceptibility. Physical desensitization solutions are limited by the scarcity of semantic information, making it difficult to cope with complex scenarios; while traditional algorithmic desensitization struggles to balance computational efficiency and privacy security, often falling into a one loses the other dilemma. Faced with this technical deadlock, there is an urgent need for a novel perception paradigm that can achieve complete decoupling of semantics-identity at the source, preserving key behavioral features while eliminating privacy backtracking from the underlying architecture. Targeting the aforementioned limitations, this paper proposes a edge-cloud collaborative \cite{yuan2025task} privacy-preserving perception technology based on the AI Flow \cite{shao2025ai,an2026ai} theoretical system. This technology aims to break the zero-sum game between security and privacy, reshaping the definition of perception in public spaces. We advocate transforming video surveillance into de-identified behavior perception, meaning the system is stripped of the capability to see privacy in its architectural design, establishing a new intelligent monitoring path that identifies only risks without involving privacy.

This study constructs core mechanisms of source desensitization \citep{liang2025integrating} and irreversible feature mapping, adopting a edge-cloud collaborative inference architecture:

\begin{itemize}
    \item \textbf{Source Desensitization and Irreversible Encoding:} In the millisecond-level instant of data acquisition, the edge device utilizes a constrained feature learning algorithm based on Information Bottleneck theory to transform raw images into abstract feature vectors. By applying non-linear mapping in high-dimensional feature space and injecting random perturbations and Gaussian noise, the system constructs a unidirectional information flow, forcibly discarding identity-sensitive information such as faces and textures.
    \item \textbf{Edge-Cloud Collaborative Inference:} The system strictly restricts raw data from leaving the edge device. The edge model is solely responsible for visual encoding and privacy stripping, transmitting the generated irreversible feature vectors to the cloud; the cloud utilizes multimodal family models \citep{song2025theoretical} to perform joint inference only on these abstract vectors, directly outputting behavioral conclusions such as falling or smoking. From both mathematical and physical perspectives, this architecture severs the path of inferring original images from the feature space, ensuring that even if data is intercepted, it is impossible to reconstruct the footage or identify individuals \citep{wu2025multi}.
\end{itemize}

The technology proposed in this study theoretically conquers the privacy-utility-efficiency impossible triangle that has long plagued the public safety field. Through the source desensitization and Edge-cloud collaborative architecture based on AI Flow, the system successfully achieves the orthogonal decoupling of semantic understanding and identity information, mathematically proving the feasibility of understanding behavior without seeing faces. This is not only a subversion of the traditional "what you see is what you get" mode of video surveillance but also provides a brand-new theoretical paradigm for the efficient deployment of privacy computing at the edge. At the application level, this solution provides a compliant and feasible technical solution for traditional security lawless zones such as restrooms, dormitories, and wards. Compared to expensive physical sensors or high-computing-power encryption schemes, the TeleAI solution achieves full coverage of high-sensitivity areas at extremely low marginal costs by leveraging existing camera infrastructure, significantly improving the refined governance level of parks and cities, truly achieving safety without dead angles, privacy without violation. More profoundly, TeleAI establishes a new ethical benchmark shifting from passive surveillance to active guardianship. It enables AI to learn see no evil, keenly perceiving risks even under the premise of invisible footage. This technical restraint and concession is precisely the greatest respect for personal dignity and privacy rights in the digital age, marking a transition in public space governance from extensive full recording to a new stage of human-centric intelligent good governance.
\section{Architecture}
\begin{figure}[h]
    \centering
    \includegraphics[width=1\linewidth]{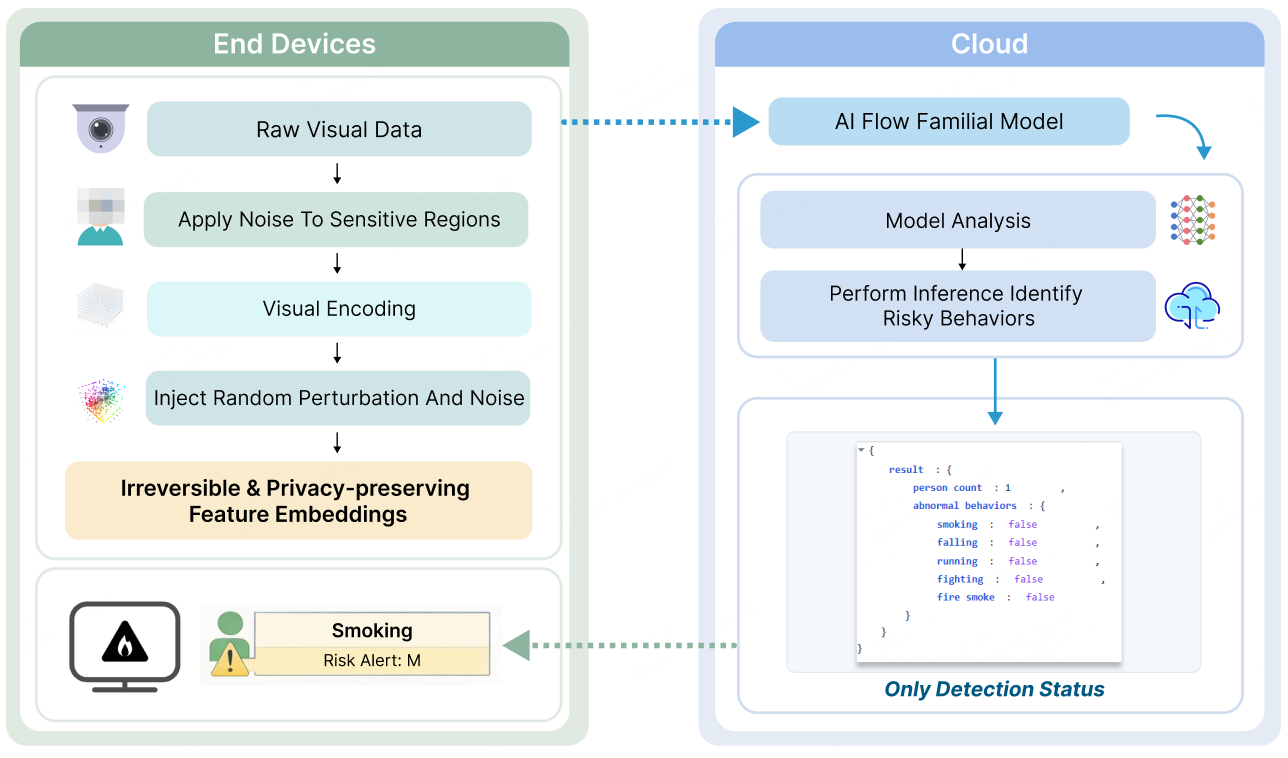}
    \caption{The architecture of the Edge-Cloud Collaborative Privacy Protection System.}
    \label{fig:placeholder1}
\end{figure}

As illustrated in the Figure \ref{fig:placeholder1}, this system constructs a Edge-cloud collaborative real-time privacy protection framework. At the edge (End Devices), the system first injects structured adversarial perturbations into Privacy-Sensitive Zones (PSZ) to discard sensitive information strongly correlated with identity. Simultaneously, it transforms concrete video streams into Irreversible Feature Embeddings, followed by the injection of random perturbations and noise to sever any path for image reconstruction. These de-identified abstract features are encrypted and transmitted to the Cloud, where they are analyzed by the AI Flow family models. The system outputs only structured text data regarding risk states—such as the detected person count and abnormal behavior status shown in the JSON example—thereby achieving precise risk recognition in private spaces without storing video or identifying individuals throughout the entire pipeline. 
\section{SPA-D Desensitization Algorithm}
Targeting the surveillance requirements of high-sensitivity environments such as restrooms, this system proposes a source desensitization technology termed SPA-D, standing for Selective Privacy-Attention Decoupling. Distinct from conventional full-image blurring or mosaic masking, SPA-D delves into the feature extraction mechanisms of Vision-Language Models (VLMs). Drawing inspiration from the VIP framework \citep{meftah2025vip}, we implement a strategy where by injecting specific structured micro-perturbations at the edge imaging source, the system achieves a mathematical erasure of Privacy-Sensitive Zones (PSZ) within the image. This method preserves global image semantics, such as characteristics of falling or fighting, while effectively blocking the model's attentional focus on PSZ areas like facial and somatic features, thereby constructing a unidirectional and irreversible information flow. 
\subsection{Privacy Adversarial Game Model}
Under the SPA-D framework, we formulate privacy protection as a constrained adversarial optimization problem. The system aims to identify an optimal image perturbation $\delta$ such that the generated desensitized image $x_{safe}$ successfully deceives the model's privacy recognition mechanism while retaining sensitivity to safety risk events. Let $x$ denote the raw image, $\mathcal{R}$ be the set of descriptors for Privacy-Sensitive Zones (PSZ) (e.g., face, facial features), and $\mathcal{T}$ be the set of public safety risk descriptors (e.g., smoking, violent physical conflict). Our optimization objective function $\mathcal{O}$ is defined as follows:

\begin{equation}
    \min_{\delta} \mathcal{L}_{sem}(\mathcal{M}(x, t), \mathcal{M}(x_{safe}, t)) - \lambda \cdot \mathcal{L}_{PSZ}(\mathcal{M}(x_{safe}, r))
\end{equation}

where $x_{safe} = x + \delta$ represents the image after source desensitization, $\mathcal{L}_{sem}$ quantifies the semantic consistency of ``safety behaviors'' between the original and desensitized images, $\mathcal{L}_{PSZ}$ measures the degree of correlation between the image and the features of ``Privacy-Sensitive Zones'', and $\lambda$ acts as a balancing coefficient used to modulate the intensity of privacy stripping. By solving this equation, we mathematically guarantee an optimal equilibrium between maximizing risk recognition capability and minimizing privacy leakage risk.
\subsection{Dual-Feature Blocking Targeting PSZ}
\subsubsection{PSZ Attention Redirection}
Vision Transformer models rely on the Self-Attention mechanism to aggregate image information. For the $h$-th attention head in the $l$-th layer, its attention map $\mathcal{A}$ determines where the model looks.
The SPA-D algorithm forcibly minimizes the weight response of the Privacy-Sensitive Zone (PSZ) in the attention map by iteratively optimizing the perturbation $\delta$. The mathematical constraint is:

\begin{equation}
    \mathcal{L}_{att} = \sum_{l=1}^{L_{depth}} \sum_{h=1}^{H} \sum_{j \in \mathcal{S}_{PSZ}} \mathcal{A}^{(l,h)}(\cdot, j) \rightarrow 0
\end{equation}

Where $\mathcal{S}_{PSZ}$ represents the set of indices for image patches located within the privacy-sensitive zone. This process is equivalent to putting selective blinders on the AI model, causing its gaze to automatically avoid PSZ regions (such as faces), thereby preventing the aggregation of private pixels into identifiable features.
\subsubsection{Sensitive Value Suppression}

To further prevent private information leakage through residual features, SPA-D introduces a Value Matrix ($V$) suppression mechanism. The Value Matrix carries the specific content information of the image. We directly weaken its information expression intensity by minimizing the $L_2$ norm of the Tokens corresponding to the PSZ:

\begin{equation}
    \mathcal{L}_{val} = \sum_{j \in \mathcal{S}_{PSZ}} ||V^{(l,h)}(j)||_2
\end{equation}

By jointly optimizing the two objectives above, SPA-D can compress the information density of the Privacy-Sensitive Zone to near zero, making it impossible to reconstruct the original image even if an attacker intercepts the feature data.
By solving this equation, we mathematically guarantee an optimal equilibrium between maximizing risk recognition capability and minimizing privacy leakage risk.

\subsection*{Dynamic Noise \& Edge-Cloud Collaboration}

TeleAI privacy protection system's noise injection process is not a simple superposition of random noise, but a precise feature interference process based on gradients. The specific implementation steps are as follows:

\textbf{Edge-Side PSZ Locking:} The lightweight model on the edge side detects the screen in real-time, quickly locating non-public Privacy-Sensitive Zones ($\mathcal{S}_{PSZ}$), such as faces and private parts.

\textbf{Gradient-Guided Noise Injection:} The system computes the gradient $\nabla_{\delta}\mathcal{L}$ based on the loss function from Section 3.2 and generates the adversarial perturbation $\delta$. This perturbation appears as Gaussian-like noise textures in the pixel space, but acts as a precise firewall targeting the PSZ in the feature space. The update formula is:

\begin{equation}
\delta_{t+1} = \delta_{t} - \alpha \cdot \text{Sign}(\nabla_{\delta_{t}}(\mathcal{L}_{att} + \lambda_{v}\mathcal{L}_{val}))
\end{equation}

\textbf{Unidirectional Information Flow Construction:} The noise-injected image $x_{safe}$ is immediately converted into abstract feature vectors on the edge side. Due to the non-linear mapping effect of the perturbation $\delta$, this conversion process loses key high-frequency information required to reconstruct PSZ details.

\textbf{Cloud-Side Secure Inference:} The desensitized feature stream is uploaded to the homologous family model on the cloud. Since SPA-D only suppresses the features of the Privacy-Sensitive Zones, the cloud model can still clearly understand limb movements and environmental states, thereby accurately identifying risky behaviors such as smoking or school bullying.
\subsection{System Implementation and Demonstration}

\begin{figure}[h]
    \centering
    \begin{minipage}[t]{0.385\linewidth}
        \centering
        \includegraphics[width=\linewidth]{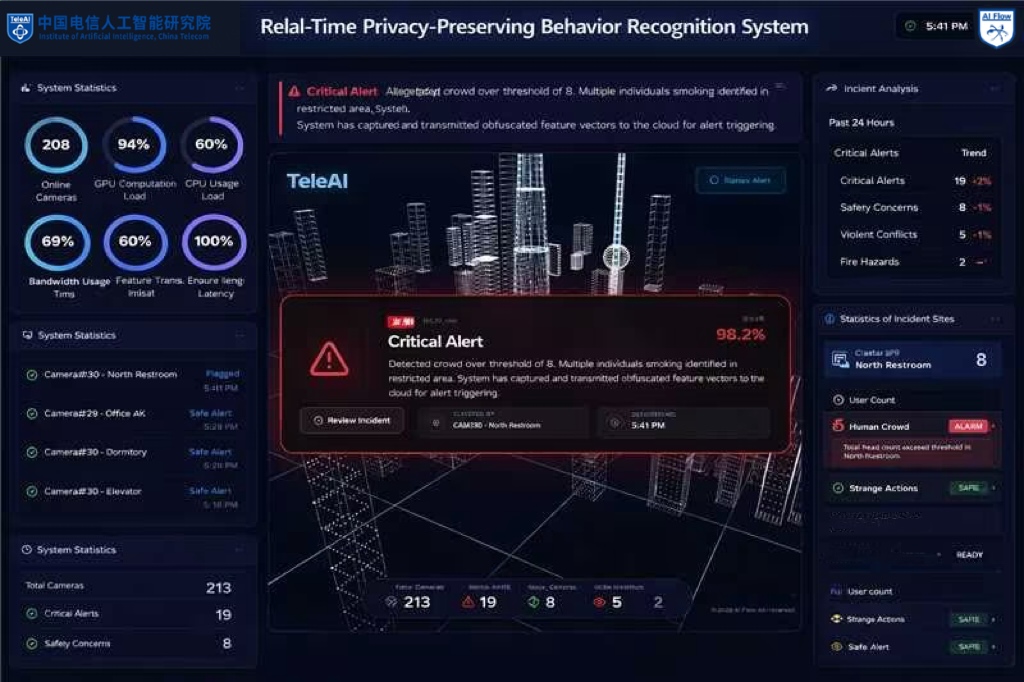}
        \caption{\textbf{Visualization of the TeleAI Management Dashboard.} 
        The interface visualizes real-time analytics including terminal load, behavior risk statistics (e.g., smoking, intrusion), and safety alerts. Note that the system reports a ``Violation Alert'' for smoking solely through text and icon indicators, with no visual access to the actual scene, verifying the privacy-first design.}
        \label{fig:dashboard1}
    \end{minipage}
    \hfill
    \begin{minipage}[t]{0.58\linewidth}
        \centering
        \includegraphics[width=\linewidth]{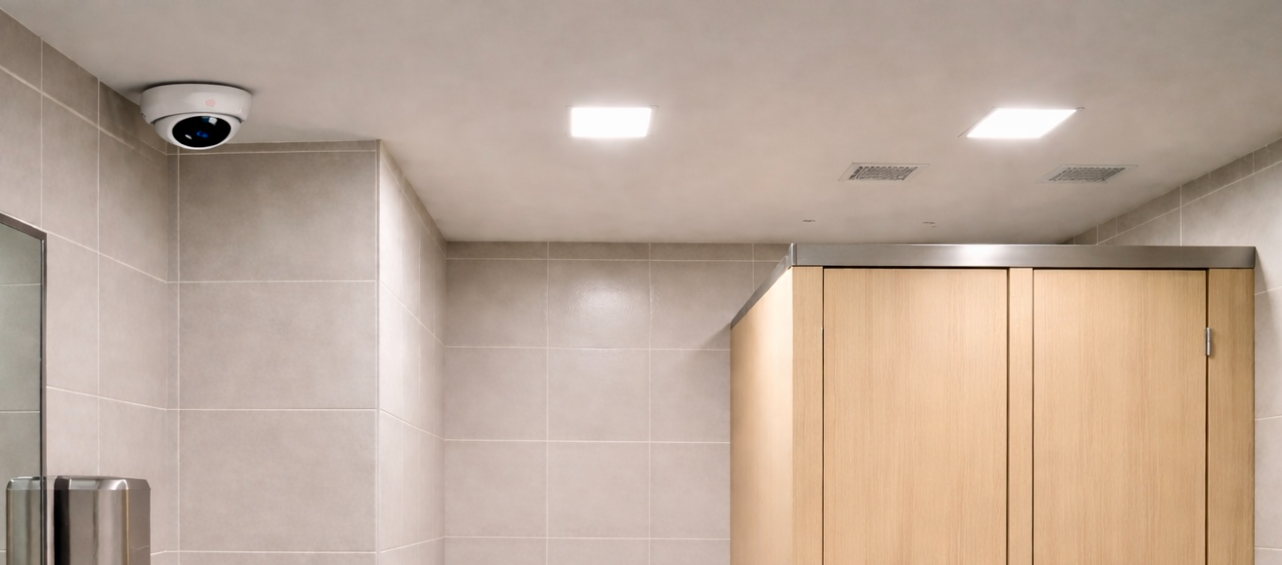}
        \caption{\textbf{Hardware Deployment in High-Sensitivity Scenarios.} 
        The TeleAI edge device installed on the ceiling of a public restroom. This node executes the source-side irreversible feature mapping, ensuring that raw visual data never leaves the local environment.}
        \label{fig:hardware2}
    \end{minipage}
\end{figure}

To validate the feasibility of the proposed Edge-cloud collaborative architecture, we deployed the TeleAI prototype in a real-world high-sensitivity environment. As shown in Figure \ref{fig:dashboard1}, the edge perception node is installed in a public restroom facility. While physically resembling a standard dome camera, the device is embedded with the SPA-D algorithm to perform millisecond-level source desensitization.

Figure \ref{fig:hardware2} presents the centralized management dashboard of the AI Flow Privacy-Preserving Real-Time Protection System. The interface demonstrates the system's capability to monitor complex risk behaviors in real-time. Notably, the dashboard strictly displays structured metadata (e.g., crowd density, specific behavioral alarms like ``Smoking Detected'') and abstract 3D spatial representations. The absence of any live video feed or identifiable facial features in the monitoring interface confirms that our system successfully achieves the ``semantics-identity'' decoupling, enabling precise risk governance without compromising user privacy.
\clearpage
\section{Conclusion}
Addressing the privacy-security paradox in public space perception, this paper proposes a new Edge-cloud collaborative paradigm based on the AI Flow framework. By introducing the SPA-D source desensitization mechanism, we achieve the orthogonal decoupling of semantic understanding and identity information at the mathematical level, breaking the long-standing zero-sum game. This solution not only theoretically conquers the "privacy-utility-efficiency" impossible triangle but also provides a compliant and feasible technical path for intelligent governance in high-sensitivity areas such as restrooms and wards. Currently, this work is under continuous refinement. Future work will further explore stronger formal privacy guarantee theories, the stability analysis of the closed-loop real-time protection mechanism, and generalization capabilities across diverse scenarios and facilities. Furthermore, we aim to achieve deep integration with regulatory and ethical compliance frameworks to drive the large-scale deployment of the TeleAI Privacy-Preserving Real-Time Protection System in broader real-world scenarios. 

\bibliographystyle{plainnat}
\bibliography{paper}

\end{document}